\documentclass[12pt]{lodziopart}          
\newcommand{\eqb}{\begin{eqnarray}}
\newcommand{\eqe}{\end{eqnarray}}
\newcommand{\diff}{{\rm d}}
\newcommand{\kappaperp}{\kappa_{\scriptscriptstyle\bot}}
\newcommand{\kappapar}{\kappa_{\scriptscriptstyle\|}}
\newcommand{\kappabohm}{\kappa_{\rm Bohm}}

\newcommand{\lesim}{\,\raisebox{-0.4ex}{$\stackrel{<}{\scriptstyle\sim}$}\,}
\newcommand{\tacc}{t_{\rm acc}}
\begin{document}
\title{Shock Acceleration of Cosmic Rays - a critical  review}
\author{J.G. Kirk\dag\ and R.O. Dendy\ddag}
\ead{John.Kirk@mpi-hd.mpg.de}
\address{\dag\ Max-Planck-Institut f\"ur Kernphysik, Postfach 10 39 80, 
D--69029 Heidelberg, Germany}

\address{\ddag\ UKAEA Fusion, Culham Science Centre, Abingdon OX14 3DB, UK} 

\begin{abstract}
Motivated by recent unsuccessful efforts to detect the predicted flux of 
TeV gamma-rays from supernova remnants, 
we present a critical examination of 
the theory on which these predictions are based. 
Three crucial problems are identified: injection, maximum achievable
particle energy 
and spectral index. In each case significant new 
advances in understanding have been achieved, which cast doubt on
prevailing paradigms such as Bohm diffusion and single-fluid MHD. This 
indicates that
more realistic analytical models, backed by more
sophisticated numerical techniques should be employed to
obtain reliable predictions. Preliminary work on incorporating the effects of 
anomalous transport suggest that the resulting spectrum should be significantly
softer than that predicted by conventional theory.
\end{abstract}

\section{Introduction}
Supernova remnants (SNR) are the favourite candidates for the source of 
\lq galactic\rq\  cosmic rays (e.g., \cite{berezinskiietal90}) and 
the best way to test this hypothesis 
is to search for the gamma-ray signal \cite{aharonian99}. In the case
of hadrons, this is produced 
by the pions from nuclear collisions; in the case of 
leptons, by bremsstrahlung or by 
inverse Compton scattering of the cosmic microwave background radiation.  
Such a signal 
should ideally 
indicate that the nuclear component of 
cosmic rays in SNR is much younger than that in 
the interstellar medium, or, failing this, that it is much more
intense in SNR than in the interstellar medium. 
To date, no unambiguous signal
has been detected which fulfils either of these criteria.

In order to predict the signal, it is necessary to propose a detailed theory
of cosmic ray acceleration, which inevitably involves
many simplifying assumptions. Using as a starting point the mechanism of 
diffusive shock acceleration (for a review see \cite{drury83}), this 
task has been undertaken by several groups 
(e.g., 
\cite{druryetal94,naitotakahara94,berezhkovoelk97,mastichiadis96,
pohl96,sturneretal97,berezhkovoelk00,ellisonetal00}).
The predicted flux of TeV gamma-rays produced by the {\em hadronic} 
component of
cosmic rays from a number of nearby SNR 
lies near the sensitivity 
limit of present day imaging \v{C}erenkov telescopes. However, the only
detections reported to date are of objects where energetic 
{\em leptons} are at least equally plausible as radiating particles
\cite{aharonian99}. Those objects in which hadrons
are expected to dominate 
do not show emission at the level hoped for
\cite{buckleyetal98,druryetal94,aharonianetal94}. In the case of Tycho's supernova remnant
the observational upper limit is 
very close to a recent, conservative prediction of the flux 
\cite{rowelletal00,voelk97}.

The indirect arguments in favour of the origin of (hadronic) cosmic rays in
SNR (notably the energy budget \cite{axford81})
remain reasonably convincing. Therefore, 
in view of the observational situation, it seems
appropriate to re-examine the status of the theoretical arguments used to 
predict the gamma-ray fluxes. In this paper we concentrate on 
recent progress pertaining to three aspects:
the injection problem, the question of the maximum achievable particle energy,
and the expected spectral slope of accelerated particles. 
 
\section{The injection problem}
The fundamental assumption of the theory of diffusive shock acceleration is
that accelerated particles diffuse in space --- i.e., that the particle
flux is proportional to the gradient of the particle density (Fick's law).
Charged particles deflected by fluctuations in the electromagnetic
fields obey this relation only if their velocities are distributed almost
isotropically. More precisely, the theory employs an expansion in
the ratio of the plasma speed in the shock frame to 
the particle speed, and the velocity
anisotropy is taken to be first order in this small parameter
\cite{drury83}. At a shock front, the
downstream plasma speed is of the same order as the thermal speed of the
ions in the plasma, so that the theory does not apply to particles whose
energy is less than several times the thermal energy. The question of how
particles might be accelerated from the thermal pool up to an energy where 
they can be assumed to diffuse is referred to as the 
\lq injection problem\rq, and cannot be treated within the framework of the
diffusive acceleration theory.

Nevertheless, this is an important question both for cosmic ray composition
and for the efficiency with which the acceleration process can proceed.
In the case of ions, much numerical work has been performed with Monte-Carlo
simulations and hybrid codes -- for a review see \cite{scholeretal98}. 
Although it is now generally accepted that the collision operator used by
early Monte-Carlo simulations (e.g., \cite{ellisonbaringjones95})
is too simple to describe injection, this technique has the advantage of 
being tractable in three dimensions. In contrast, hybrid simulations of 
the shock structure which directly show the ion population emerging from the 
thermal pool are restricted to two space dimensions. Consequently,
cross-field transport is not treated consistently
\cite{giacalonejokipiikota94}, which might 
be important for the injection mechanism \cite{giacaloneellison00}. 

In addition to numerical work, an analytic theory of injection has been
developed by Malkov \& V\"olk \cite{malkovvoelk95}. In this, a
thermostat model of the shock is employed in which some fraction of the
hot downstream particles is assumed to stream into the upstream plasma.  
The resulting ion beam excites Alfv\'en waves via the cyclotron 
instability, which scatter the particles in pitch-angle. Thus, the physical
ingredients are the same as
in standard cosmic ray transport theory. However, because of the inherent
anisotropy of the mildly suprathermal particles, it is necessary to adopt 
a more refined treatment of the transport \cite{kirkschneider89}.
To complement this model, Malkov \cite{malkov97a} 
has developed a theory of the 
thermostat which involves a large amplitude, monochromatic Alfv\'en wave,
whose role is to confine the majority of hot ions in the downstream region,
whilst allowing a small fraction to counter\-stream into the upstream plasma.
This theory has the advantage of offering a prescription suitable for
incorporation into combined cosmic ray and hydrodynamic codes
\cite{gieseleretal99}. However, the theory contains many simplifications 
whose validity can be tested only by extensive numerical simulation.

The problem of electron injection has long been considered to be 
even more difficult
than that of ion injection. This is because the cyclotron resonance condition
which enables ions to excite weakly damped Alfv\'en waves cannot be satisfied
by mildly supra-thermal electrons (see, for example, \cite{malkovvoelk95}).
Indeed, an electron velocity exceeding roughly 2000 times 
the Alfv\'en speed would be required for resonance, which renders the 
process irrelevant for the problem of injection.  
A viable alternative has recently been proposed by 
Dieckmann et al~\cite{dieckmannetal00},
building on suggestions by Galeev \cite{galeev84}
and Galeev et al~\cite{galeevetal95} and by 
McClements et al~\cite{mcclementsetal97}. Dieckmann et~al show that
shock-reflected ion populations in the upstream plasma can drive collective
instabilities, such that the resultant waves excited in the plasma damp on
thermal electrons, thereby accelerating them across the magnetic field to
mildly relativistic energies. A fully self-consistent treatment of this
fundamentally nonlinear plasma process is obtained using large-scale
particle-in-cell simulations, and there is close quantitative agreement with
analytical theory where points of contact exist. The ion population parameters
are initialised on the basis of shock acceleration models and parameters, for
example those of Cargill \& Papadopoulos \cite{cargillpapadopoulos88}.
Thereafter, broadband electrostatic field oscillations grow in the frequency
range between the electron plasma- and gyro-frequencies, excited initially by 
Buneman-type instabilities, with episodes of high field temporaly correlated
with episodes of electron energisation. Several resonant and non-resonant
mechanisms for the latter appear to be at work, of which the strongest
involves stochastic wave-particle interactions. Dieckman et al demonstrate
conclusively that this \lq\lq bootstrap\rq\rq\ mechanism can energise electron
populations from background to characteristic energies above $10\,$keV, 
(sufficient to account for the hard X-ray emission from, for example, 
Cas-A \cite{laming00} or from clusters of galaxies \cite{sarazinklempner00})
with individual electrons accelerated to
several tens of keV. Under typical conditions in the interstellar medium,
this would enable them to fulfil the cyclotron
resonance condition with weakly damped 
Alfv\'en waves.

\section{Maximum achievable energy}

To order of magnitude accuracy, the
acceleration timescale $\tacc$ at a shock front
can be estimated as
\eqb
\tacc&\approx&\kappa/u^2
\eqe
where $\kappa$ is the spatial diffusion coefficient and $u$ is the shock speed.
Unfortunately, the value of $\kappa$ is not directly measurable. Furthermore,
it depends on the particle energy (through their Larmor radius) and
the properties of the plasma turbulence
(which is probably driven by the 
accelerated particles themselves \cite{bell78}), so 
that it is difficult to estimate reliably. Nevertheless, it is usual to 
assume that
there exists a \lq self-quenching\rq\ mechanism in operation
which limits the
amplitude $\delta B$ of the turbulent fluctuations in the magnetic field $B$
near the shock front such that $|\delta B|\lesim B$ 
\cite{voelk81,mckenzievoelk82}. In this case, one is led to the
estimate
\eqb
\kappa&<&\kappabohm
\label{bohmlimit}\\
&=&
{2v^2\gamma mc\over 15ZeB}
\label{bohmvalue}
\eqe
where $v$, $\gamma$, $m$ and $Ze$ are the particle's velocity, Lorentz factor,
mass and charge and $B$ is the average magnetic field strength.

In the standard picture of cosmic ray acceleration in a supernova remnant, the 
maximum energy achieved is determined in the free expansion phase 
(e.g., \cite{lagagecesarsky83}). 
If the diffusion coefficient is limited according to
Eq.~(\ref{bohmlimit}), this phase does 
not really 
last long enough to allow cosmic rays to be accelerated up to the 
observed 
knee in the spectrum, at
$\sim10^{15}$--$10^{16}\,$eV: setting the 
acceleration timescale equal to the age of the shock front 
gives as an estimate of the maximum energy
\eqb
E_{\rm max}&=&6\times10^{13} Z\left(u\over3000\,{\rm km\,s^{-1}}\right)^2
\left({t_{\rm sw}\over 300\,{\rm yr}}\right)\left({B\over 1\mu{\rm G}}\right)
\,{\rm eV}
\label{standardestimate}
\eqe
in terms of the 
shock speed $u$ during the free-expansion phase and the time $t_{\rm sw}$
available before this phase ends. 
A more careful analysis --- amongst other effects, allowing 
for the fact that the limit given in
Eq.~(\ref{bohmlimit}) should be reached only at the shock front and not
throughout the upstream plasma ---  has been 
performed by Lagage \& Cesarsky \cite{lagagecesarsky83}, who conclude that the
Eq.~(\ref{standardestimate}) is an over estimate by at least a factor of six.

Several papers have suggested ways out of this difficulty, but, until
recently, none of these 
has seemed very promising. Jokipii \cite{jokipii87}, for example, pointed
out that, according to the quasi-linear theory of plasma turbulence, the
diffusion coefficients parallel and perpendicular to the magnetic field
satisfy
\eqb
\kappaperp&\ll&\kappabohm\,\ll\,\kappapar
\eqe
so that perpendicular shocks should accelerate particles much faster 
than parallel ones. However, to make a substantial difference for 
supernova remnant shocks,
this not only would require fine-tuning of the interstellar 
magnetic field to make
the shock front perpendicular, but would also only be effective 
if, 
contrary to expectation, the
level of turbulence stayed well below the self-quenching limit 
i.e., if $|\delta B|\ll B$. 
Using a more conventional approach,
Berezhko \cite{berezhko96} has presented an re-analysis of the
estimate (\ref{standardestimate}), arriving at a value 
$E_{\rm max}\approx 10^{15}\,$eV. However, much of the enhancement 
he found arises from the reflecting boundary condition imposed 
at the so-called \lq piston\rq. This is a numerical device used to 
drive the exploding remnant in 
a hydrodynamic simulation, and the boundary condition is 
chosen for convenience. It is not clear that there is a real
physical basis to the effects it produces. Furthermore, 
Berezhko did not take account of the reduction 
of $E_{\rm max}$ by the effects which 
Lagage \& Cesarsky \cite{lagagecesarsky83} had found to be important. 
 
However, in an interesting recent 
development, Lucek \& Bell \cite{lucekbell00} have performed numerical 
simulations of the turbulence driven by 
cosmic ray streaming at a shock front. This seems to be the only 
practicable way of advancing our understanding beyond that of
weakly nonlinear computations of growth and damping 
\cite{voelk81,mckenzievoelk82}.
In particular, it 
provides an opportunity of testing the \lq self-quenching\rq\ 
hypothesis $|\delta B|\lesim B$ \cite{voelk81}, 
which is the crucial ingredient of the limit given in Eq.~(\ref{bohmlimit}).
The simulations, which are fully 3-dimensional, use an MHD code for the 
background plasma, together with a kinetic description of the accelerated 
cosmic rays.
Initial conditions are chosen with an idealised energy 
distribution of isotropic cosmic rays, superposed on a background 
plasma which streams in the direction of the magnetic field at an Alfv\'en Mach number of 10. 
The temporal evolution is followed until the instability saturates. It is 
found that the linearly most unstable Alfv\'en wave grows with approximately 
the linear growth rate until well into the nonlinear regime. The magnetic 
field of this wave, which is directed perpendicular to the streaming 
direction, eventually dominates the initial field, causing the instability to 
saturate when it is strong enough to absorb the energy associated with the 
streaming. At this stage, $|\delta B|\gg B$, --- well above the proposed 
self-quenching value --- and an initially parallel 
shock front (which is not included explicitly in the simulations) would 
have become essentially perpendicular. 

These results have a major impact on the estimate of $E_{\rm max}$ 
described above. The strong amplification of 
the magnetic field means that in estimating the 
Bohm diffusion coefficient according to Eq.~(\ref{bohmvalue}) it is not 
realistic to insert the value of the magnetic field in the unperturbed 
interstellar medium. Even if we ignore the difference 
between a parallel and quasi-perpendicular shock, and set 
$\kappapar\sim\kappaperp\sim\kappabohm$, a substantial increase 
in $E_{\rm max}$ results. Lucek \& Bell \cite{lucekbell00} estimate a 
field enhancement of a factor of 1000 which, using the estimate of 
Lagage \& Cesarsky \cite{lagagecesarsky83} implies 
$E_{\rm max}\sim10^{16}\,$eV.

\section{Spectral index}

Adopting an arbitrary power-law distribution
for the accelerated particles, 
Gaisser et al~\cite{gaisseretal98} find that the best fit to 
the EGRET observations of the SNR IC443 and $\gamma$-Cygni
is
achieved with a cosmic ray spectrum given by 
$-\diff \ln N/\diff \ln E\equiv s=2.4$, (where $\diff N(E)$ is the 
differential number of cosmic rays 
in the remnant with energy between $E$ and $E+\diff E$) 
which cuts off at about $80\,$TeV, 
well above the EGRET energy range.
Theoretical models of nonlinear diffusive acceleration, on the other hand,
give spectra at the position of the shock front which are rather harder.
For example, both stationary Monte-Carlo spectra in plane geometry 
and time-dependent kinetic computations in spherical geometry agree on the
basic shape to be expected \cite{berezhkoellison99}, which is very hard 
($s<2$) below 
the cut-off energy. An even harder ($s=1.5$) spectrum is found in the 
full analytic solution of the stationary 
plane case \cite{malkov97b,malkovetal00} --- a difference
which might be due to the
technical details associated 
with the assumptions concerning particle escape, 
but which has not yet been investigated in depth.
The gamma-ray emission of a SNR
depends not just on the instantaneous spectrum of particles at the shock
front, but on a superposition of particle spectra throughout the remnant.
Nevertheless, the integrated particle spectrum has a slope
$s\lesim2$, which hardens towards higher energy
\cite{berezhkovoelk97,ellisonetal00} and could not produce the type of
particle spectrum favoured by Gaisser et al~\cite{gaisseretal98}.
A hard high-energy spectrum ought to be more easily seen at TeV rather 
than GeV energies, but results to date are discouraging \cite{buckleyetal98}.
Only two or, perhaps, three shell-type SNR have so far been 
detected \cite{tanimorietal98,puehlhoferetal99,muraishietal00} and it is quite 
possible that this 
emission arises from relativistic electrons
of fairly soft spectrum \cite{aharonianatoyan99,atoyanetal00}, rather than
the predicted hard-spectrum protons. 

However, not only the maximum energy, but also the spectral slope 
of cosmic rays produced by a SNR shock is strongly affected by the properties
of the self-induced turbulence if, as suggested in the previous section,
the self-quenching mechanism fails to maintain $|\delta B|\lesim B$.  
This is because the magnetic field fluctuations generated lie 
in the plane of the shock front. Acceleration of cosmic rays therefore involves
the cross-field transport properties of the plasma. 
Such a situation has already 
been investigated in the test-particle 
approximation \cite{duffyetal95}, where it was found that
an anomalous, non-diffusive, transport mode arises due to the statistical
wandering or \lq braiding\rq\ of the magnetic field lines. This 
had previously been studied in connection with the propagation of 
cosmic rays through the galactic magnetic field 
\cite{getmansev63,jokipiiparker69,chuvilginptuskin93} and is
well-known in fusion plasmas \cite{isichenko91,raxwhite92}. Its effect is to
soften the spectrum of accelerated particles \cite{kirketal96,gieselerkirk99}
--- in the case of test particles at a shock of compression ratio 4,
from the familiar result of diffusive
shock acceleration: $s=2$ to the value $s=2.5$. Physically, this arises 
because the anomalous transport mode 
inhibits cross-field propagation, and tends to sweep 
particles away from the shock front into the downstream plasma
more effectively than does diffusive transport.
No nonlinear calculations of
this process have yet been attempted, but a softer spectrum, more in line
with that suggested by the gamma-ray observations, is clearly to be expected.

\section{Conclusions}

The standard picture of the diffusive acceleration of cosmic rays 
at a supernova shock front faces several well-known and 
stubborn problems. Nevertheless, progress has been substantial, and 
new perspectives are emerging:
\begin{itemize}
\item
In the case of electron injection, large scale particle-in-cell
simulations have shown that energisation can occur in the 
turbulence driven by a population of reflected ions
\cite{dieckmannetal00}. 
For ion injection, an analytic theory is avaliable which describes the 
injection process at a parallel shock, given that some fraction of the 
thermal ions counter-stream \cite{malkovvoelk95}.
An analytic
theory is also 
under development which aims to provide an understanding of how such 
counter-streaming ions can be generated self-consistently \cite{malkov97a}.
\item
The failure of the standard 
mechanism to accelerate particles up to $10^{16}\,$eV, 
appears to be due to an underestimate of the importance of self-generated
turbulence at the shock front. Once again, this realisation 
has emerged from computer simulation work \cite{lucekbell00}.
\item
The hard spectra predicted in the standard picture, which 
have not been confirmed by observation, are also based on 
an assumption about the self-generated turbulence which is now open to 
question. The modifications which are introduced by anomalous transport 
properties have been shown to produce softer spectra, more in line with 
the constraints 
inferred from observations of gamma-rays
\cite{duffyetal95,kirketal96,gieselerkirk99}. 
\end{itemize} 

\ack
This work is a collaboration of the European Network 
{\em AstroPlasmaPhysics} supported by the European 
Commission under the TMR-programme,
contract number ERBFMRX-CT-98-0168, and was supported in part 
by the UK DTI. 
We thank Felix Aharonian and Heinz V\"olk 
for helpful comments.

\end{document}